\documentstyle[11pt,newpasp,twoside,epsf]{article}
\def\lesssim{\mathrel{\hbox{\rlap{\hbox{%
 \lower4pt\hbox{$\sim$}}}\hbox{$<$}}}}
\def\gtrsim{\mathrel{\hbox{\rlap{\hbox{%
 \lower4pt\hbox{$\sim$}}}\hbox{$>$}}}}

\def\edcomment#1{\iffalse\marginpar{\raggedright\sl#1\/}\else\relax\fi}
\reversemarginpar

\begin{document}
\title{Spectroscopic and Variability Surveys for AGN in the Groth Survey Strip}
\author{Vicki Sarajedini}
\affil{Department of Astronomy, University of Florida, Gainesville, FL 32611}

\begin{abstract}
Preliminary results are presented for a spectroscopic survey of the
Groth Survey Strip (GSS), a 40 by 3.5 arcminute region of the sky imaged with
HST, for which several hundred galaxy spectra have been obtained as part
of the DEEP project (http://deep.ucolick.org).  
At least 6 broad-line AGNs (primarily Seyfert 1s) have been detected as well
as several narrow-line Seyfert 2 candidates.  The Seyfert galaxies detected
in our survey have integrated absolute magnitudes extending to M$_B$$\sim$-17.5,
probing fainter magnitudes and higher redshifts than existing optical 
spectroscopic 
surveys.  We also discuss a variability study of the GSS using the original 
HST images from 1994 and new images obtained in 2001.  
The high resolution obtained with HST allows us to isolate and measure 
variable galactic nuclei too faint to be detected from the ground, reaching
nuclear magnitudes of M$_B$$\sim$-16 in galaxies to z$\sim$0.8.  The combination
of these techniques provides a powerful probe of the 
population of low-luminosity AGNs at moderate redshifts.
\end{abstract}

\section{Introduction}

To better understand the nature of any class of extragalactic object, an
accurate knowledge of the luminosity function (LF) over a wide range of
absolute magnitudes and covering a range of redshifts is necessary.
The AGN LF is populated by
quasars at the brighter, primarily high redshift end and
Seyfert galaxy nuclei, considered to be their intrinsically fainter
counterparts, at the low luminosity, low redshift end (Cheng et al. 1985;
Huchra \& Burg 1992; Maiolino \& Rieke 1995).  See Wizotski (2001, this volume)
for a discussion of the current state of the AGN LF.
While bright QSOs are easily observable at all redshifts, fainter Seyfert
nuclei become increasingly difficult to detect at redshifts much 
beyond the local universe. 
Understanding how the faint end of the AGN LF evolves
is of particular importance for determining the frequency and
total space density of AGNs at earlier epochs.

The purpose of this paper is to show how the increased light 
gathering capabilities of large (i.e. 10-meter) telescopes 
and the high spatial resolution achieved with space-based telescopes
allows for the detection of these lower luminosity AGN at higher
redshifts.  In Sections 2 and 3, we describe preliminary results
of a spectroscopic survey for AGN using the Keck telescope.  Section
4 outlines plans for a variability survey for AGN in this same
region of the sky using Hubble Space Telescope images separated by
7 years.  A summary of this research is presented in the final section.

\section{The DEEP Project}
 
The Deep Extragalactic Evolutionary Probe (DEEP; Koo et al. 1996;
Simard et al. 2001) is a project
designed to study the formation and evolution of distant field galaxies
by combining images from HST with spectroscopic data from the Keck
telescope.  The Groth Survey Strip (Groth et al. 1994) is one of the 
fields targeted by DEEP for spectroscopic follow-up.
The GSS is comprised of 28 contiguous WFPC2 fields located
at 14h17m+52 imaged in the F606W and F814W filters.  
Optical spectra have been obtained through multi-slit masks
with the Keck/Low Resolution Imaging Spectrograph for 775 objects
in the GSS between 1995 and 1999.  Of the 683 spectra with high
enough S/N to identify spectral features and determine redshifts, 
634 are galaxies and 49 are galactic stars.  The galaxies
extend to I$_{AB}$$\sim$24 with a mean redshift of z$\simeq$0.8.  The typical
exposure time is one hour with some of the fainter targets being
exposed for up to a few hours.  Since the majority of galaxies at
these redshifts have sizes comparable to the seeing resolution 
($\sim$1$\arcsec$), spatial spectral information is not available in most
cases and a one-dimensional spectrum has been produced for each object
by summing several pixels along the spatial axis.

\section{Spectroscopic Identification of AGN}

Active Galactic Nuclei can be detected in galaxy spectra through the
presence of broad, permitted lines covering a wide range of ionization.
Optical spectra for the galaxies studied in this survey display 
broad H balmer series lines or singly ionized MgII emission (at rest 
$\lambda$2800$\AA$ for higher z objects) when a Seyfert 1 nucleus
or QSO is present in the galaxy.  Narrow-line Seyfert 2 or LINER galaxies
are defined by the property that their forbidden lines and permitted
lines have similar, narrow widths with line ratios indicating a more
energetic ionizing source than hot stars alone can provide (although
in the case of LINERs, this is still under discussion (see Ho 2001, 
this volume)).

At the present date, a subsample of 235 spectra from the GSS
spectroscopic survey (those obtained
between 1995 and 1997 with high enough S/N to discern spectral
features) have been analyzed to search for AGN. 
Of these objects, 7 are galactic stars and 7 additional galaxies
are high-z galaxies (z$\gtrsim$2.8) not considered in this
analysis for the sake of sample uniformity.  

Inspection of the remaining 221 galaxy spectra has 
revealed only two objects with broad lines representing 
$\sim$1\% of the galaxies.  These galaxies have redshifts of
1.15 and 1.22, each displaying broad MgII emission.
Initial inspection of the most recently obtained spectra has
revealed at least 4 more broad-line galaxies with redshifts ranging
between 0.6 and 1.  The inclusion of the additional spectra are 
consistent with the finding that
$\sim$1\% of the GSS galaxies appear to be Seyfert 1/QSOs.   
The integrated luminosities for these galaxies range from 
--19.5$\gtrsim$M$_B$$\gtrsim$--22.8, just below the nominal dividing line 
between QSOs and Seyferts.  Since these are integrated magnitudes,
the actual nuclear magnitudes are likely to be even fainter. 

The majority of the galaxies display narrow emission lines.
AGN can be differentiated from star-forming galaxies based on the
emission-line ratios of the most prominent optical lines such as
[OII]$\lambda$3727$\AA$, [OIII]$\lambda$4959,5007$\AA$,[NII]$\lambda$6548$\AA$,
[SII]$\lambda$6717,6730$\AA$, H$\alpha$ and H$\beta$ (e.g. Veilleux \&
Osterbrock 1987).  For a large fraction of galaxies in the GSS, however,
many of these lines are redshifted out of the optical range.  In addition,
our spectra, like many large, multi-slit spectroscopic surveys, are 
not flux calibrated.  For these reasons, traditional
line ratio diagnostics are not applicable to our survey.

A new emission line diagnostic (Rola, Terlevich \& Terlevich 1996) is being
employed to differentiate AGN and star-forming galaxies in our
survey spectra.  This technique is based only on the equivalent widths
of [OII] and H$\beta$, allowing for classification of galaxies to
z$\simeq$0.8 with optical spectra and avoiding the necessity of 
flux calibration.  Two distinct zones define the AGN region of the diagram,
at EW(H$\beta$)$<$10 and EW(OII)/EW(H$\beta$)$>$3.5.  Using a sample
of local emission line galaxies, Rola et al. (1996) find that 87\% of 
the AGN reside in these regions with 88\% of the HII galaxies falling
in the remaining region.  Although this technique does not perfectly
separate the two object classes, it does a fairly good job of 
identifying the majority of AGNs in a sample of emission line galaxies.

Of the 221 galaxy spectra in our subsample, 90 have both
the [OII] and H$\beta$ lines in the optical spectrum range.
Out of those 90 galaxies, only 44 show both lines in emission.
We plot these on the Rola diagram in Figure 1 with appropriate error bars.
We note the importance of correcting the H$\beta$ EW measurements
for the underlying stellar continuum absorption.  
Without the detection of the H$\alpha$ emission line 
for these galaxies, we can only estimate the amount of H$\beta$ absorption
in the continuum.  We have chosen a moderate value of
3 $\AA$ for the underlying stellar absorption (Kennicutt et al. 1992,
Tresse et al. 1996).  This correction has the
effect of pushing objects into the HII region of the diagram.
The filled circles represent those galaxies clearly in the
HII galaxy region.  The open circles are those in the AGN region but with 
error bars extending into the HII region.  The asterisks are those galaxies
clearly in the AGN region of the diagram.  If the probability that these 
galaxies are AGN is 88\%, our lower-limit estimate on the total number
of narrow-line AGN in the GSS out to z$\simeq$0.8 is 10\%.  If we
include the additional 11 objects which lie in the AGN region but
have error bars allowing them to be placed in either the AGN or HII 
regions, our fraction increases to 20\%.  

\begin{figure}[t]
\plotone{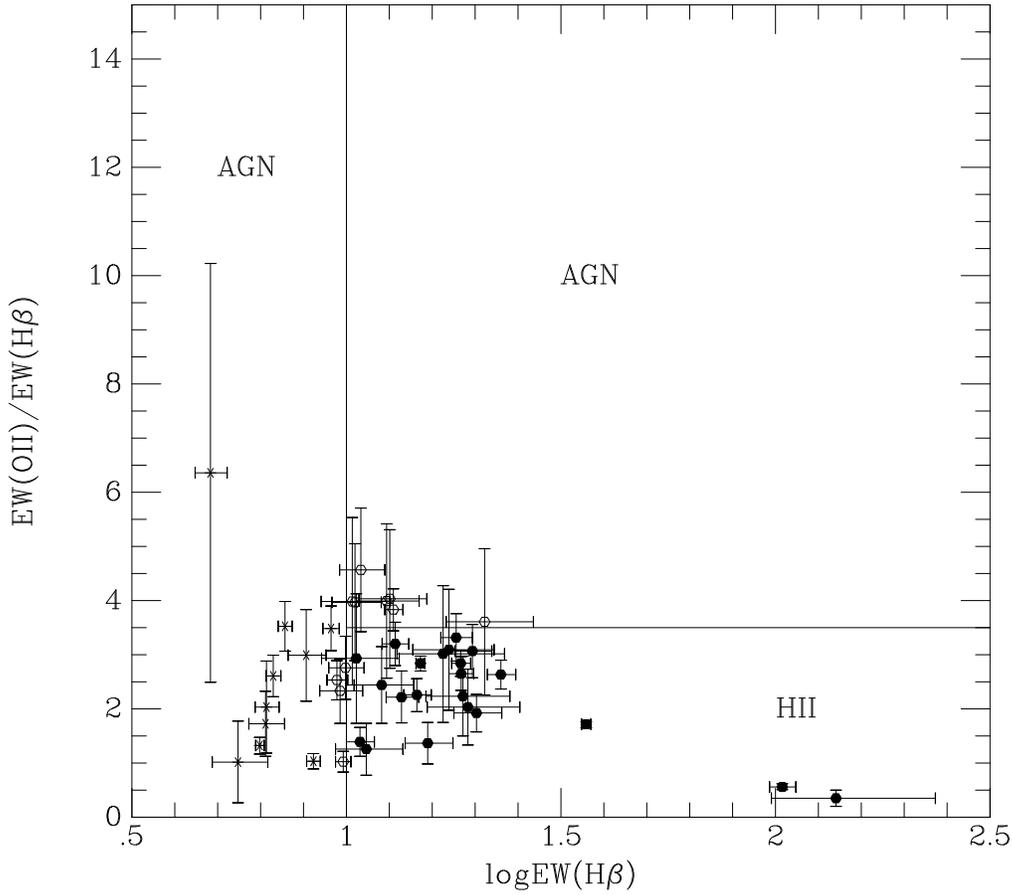}
\caption{EW[OII]/EW(H$\beta$) ratio versus EW(H$\beta$).  Solid
symbols are those galaxies in the HII region of
the diagram.  Open symbols are those in the AGN region with
error bars extending into the HII region and asterisks are
galaxies in the AGN region of the diagram.}  
\end{figure}

The AGN candidates in our survey have integrated absolute magnitudes extending
to M$_B$$\simeq$--17.5, with nuclei that may be up to a
magnitude fainter.  This demonstrates the strength of the Keck telescope
in probing activity in galaxies at much fainter luminosities and
higher redshifts than previously possible.  Our completed survey for
AGN in the GSS will extend the AGN LF several magnitudes
fainter at z$\sim$0.8.

A few other diagnostics are available to identify narrow-line 
AGN in our survey such as the strength of [NeIII]$\lambda$3869 and
the presence of [NeV]$\lambda$3426.  A few additional 
candidates have been detected
through the presence of these emission lines.
A full analysis of the complete sample of GSS galaxy spectra to
detect AGN is presented in a future paper (Sarajedini et al. 2002).

\section{Detecting AGN through Variability Surveys}

Obtaining spectra for as many galaxies as possible in a 
particular region of the sky is a robust way to find and classify
the population of active galaxies.  However, this technique requires
many nights of observing on large telescopes.  
A less ``expensive" way to detect 
AGN, and complement the spectroscopic survey, 
is through the use of multi-epoch images to identify variable sources.

Variability has long been known as an effective way to identify
QSOs (e.g. Hawkins 1986) with Koo, Kron \& Cudworth (1986)
finding $\sim$80\% of their spectroscopic and color
selected quasars in Selected Area 57 to be variable over an 11 year
time period.  In addition to quasars, Bershady et al. (1998)
detected 14 extended variable objects in this region with
Seyfert-like spectral characteristics.  The variability
amplitude for objects in SA57 was generally higher for 
active nuclei of lower luminosity,
making this technique well suited for the selection
of intrinsically faint QSOs and Seyfert-like nuclei.

We are conducting a survey to detect nuclear variabilty for galaxies
in the GSS.  In addition to the original HST images taken in 1994,
a second epoch has been obtained for the entire region in the 
spring/summer of 2001 (J. Mould, PI).          
The unique high resolution capabilities of HST are necessary
to isolate and measure faint, variable nuclei within brighter host galaxies.   
We can easily detect and
measure structural parameters for galaxies to V$_{606}$$\simeq$25
with redshifts extending to
z$\simeq$1 in the GSS (Simard et al. 1999).
Many galaxies in this regime have shown evidence of central point source
components in addition to a disk and/or bulge component
(Sarajedini et al. 1999).
Assuming a typical disk host galaxy with a scale length of 
0.25$\arcsec$, we estimate that nuclei comprising
as little as 15\% of the host galaxy light can be detected
in galaxies down to V$_{606}$$\simeq$23.5.
The advantage of HST is the ability to do accurate photometry within
smaller apertures, thus allowing us to probe much lower AGN/host galaxy
luminosity ratios than can be done from the ground.

The success of this technique has been demonstrated with the Hubble Deep Field
North.  Based on observations of the
HDF-N separated by two years, we have detected nuclear variability at or
above the 3$\sigma$ level in 8
of 633 galaxies at I$_{814}$$\leq$27.5 (Sarajedini et al. 2000).
Only 2 detections would be expected by chance in a normal distribution.
At least one of these 8 has
been spectroscopically confirmed as a Seyfert 1 galaxy.
Based on the AGN structure function for variability
(Trevese \& Kron 1990), the estimated luminosities
for the varying nuclear components extend to M$_B$$\simeq$--16 
providing an interesting comparison with
the population of local Seyfert galaxies at similar luminosities. 

These results demonstrate the strength of this technique
in probing faint AGN at z$\simeq$1. 
Extending this search to the GSS should produce a 
much larger sample of faint AGN.
Based on several different estimates for the number of AGN
in the local universe 
(e.g. Huchra \& Burg 1992; Maiolino \& Rieke 1995; Ho et al. 1997)
we expect to find at least $\sim$45 Seyfert-like nuclei in the
entire GSS (assuming no evolution to z=1) or $\sim$120
if mild number density evolution has occurred.
With this much larger sample, 
we will have the ability to not only determine number density
evolution with statistical significance, but also study any changes with
redshift in the shape of the LF for low-luminosity AGN.
 
\section{Summary and Conclusions}

With the aim of studying the evolution of AGN in the low-luminosity 
regime, we are conducting two different yet complementary
surveys of galaxies in the GSS.  We present the preliminary results
of our spectroscopic search to identify broad and narrow-line AGN
based on their spectral characteristics.  Six broad-line AGN
have currently been detected, representing about 1\% of the galaxies
for which spectra have been obtained as part of the DEEP project.
A larger number (10 to $\sim$20) of narrow-line AGN candidates have been
detected using a new emission line diagnostic to identify Seyfert 2s
and LINERs based on the equivalent widths of [OII] and H$\beta$.
This represents between 10 and 20\% of the galaxies to z$\simeq$0.8
in which these lines could be detected.  The absolute magnitudes
of these galaxies extend to M$_B$$\simeq$--17.5 with a mean
redshift of $\sim$0.8.

We have also outlined a program to search for variable nuclei in GSS
galaxies using HST images separated by 7 years.  The high resolution
capabilities of HST will allow us to detect and measure faint
nuclei using small aperture photometry, probing lower nuclear/host
galaxy ratios than possible from the ground.  Using this
technique on HDF-N images separated by 2 years, we have shown that
varying nuclei as faint as M$_B$$\simeq$--16 may be detected for
galaxies out to z$\simeq$0.8.

The results of these surveys will be used in conjunction with
GSS observations in other wavelengths such as X-ray (XMM; 
Griffiths et al. 2000), infrared (SIRTF), radio (VLA FIRST) 
and sub-millimeter (SCUBA)
to better understand the nature of the AGN and their host galaxies.

\acknowledgements
VS acknowledges the many members of the DEEP Team
(UC Santa Cruz) for work in obtaining and reducing spectral data
presented here.  Financial support for part of this work comes 
from the NSF through grant AST-9529098 and 
from NASA grants
through the Space Telescope Science Institute, operated by AURA, Inc.
under NASA contract NAS5-26555.


\begin{references}
\reference{Bershady, M. A., Trevese, D. \& Kron, R. G. 1998, ApJ, 496, 103}
\reference{Cheng, F. Z., Danese, L., De Zotti, G. \& Franchesini, A.
1985, \mnras, 212, 857}
\reference{Griffiths R. E., Miyaji, T. \& Ptak, A. 2000, BAAS 197, 5606}
\reference{Groth, E. J., Kristian, J. A., Lynds, R., O'Neil, E. J., 
Balsano, R., Rhodes, J., \& the WFPC-1 IDT.  1994, BAAS, 185, 5309}
\reference{Hawkins, M. R. S. 1986, MNRAS, 219, 417}
\reference{Ho, L. C., 2001, this volume}
\reference{Ho, L. C., Filippenko, A. \& Sargent, W. L. W. 1997, ApJS, 112, 315}
\reference{Huchra, J. \& Burg, R. 1992, ApJ, 393, 90}
\reference{Kennicutt, R. C. 1992, \apjs, 79, 255}
\reference{Koo, D. C. et al. 1996, ApJ, 469, 535}
\reference{Koo, D. C., Kron, R. G. \& Cudworth, K. M. 1986, PASP, 98, 285}
\reference{Maiolino, R. \& Rieke, G. H. 1995, ApJ, 454, 95}
\reference{Rola, C. S., Terlevich, E. \& Terlevich, R. J. 1996, \mnras, 
289, 419}
\reference{Sarajedini, V. L., Green, R. F., Griffiths, R. E., \& Ratnatunga, K.
1999, ApJ, 514, 746} 
\reference{Sarajedini, V. L., Gilliland, R. L. \& Phillips, M. M. 2000, 
AJ, 120, 2825}
\reference{Sarajedini, V. L. et al. 2002, in preparation}
\reference{Simard, L., Koo, D. C., Faber, S. M., Sarajedini, V. L.,
Vogt, N. P., Phillips, A. C., Gebhardt, K., Illingworth, G. D., \&
Wu, K. L. 1999, ApJ, 519, 563}  
\reference{Simard, L. et al. 2001, ApJ, submitted}
\reference{Tresse, L., Rola, C. S., Hammer, F., Stasinska, G.,
Le Fevre, O., Lilly, S. J. \& Crampton, D. 1996, \mnras, 281, 847}
\reference{Trevese, D. \& Kron, R. G. 1990, in Variability of Active
Galactic Nuclei, ed. H.R. Miller \& J.P. Witta (Cambridge: Cambridge
University Press), 72}
\reference{Veilleux, S. \& Osterbrock, D. E. 1987, \apjs, 63, 295}
\reference{Wisotzki, L. 2001, this volume}
\end{references}
\end{document}